\documentclass[12pt]{iopart}
\pdfoutput=1

\usepackage{graphicx}
\usepackage{color}
\begin{document}

\title[Kate Jones]{Transfer reaction experiments with radioactive beams: from halos to the r-process}

\author{Kate L. Jones}

\address{Department of Physics and Astronomy, University of Tennessee, Knoxville, TN 37996, USA}
\address{Physics Division, Oak Ridge National Laboratory, Oak Ridge, TN 37831, USA}
\ead{kgrzywac@utk.edu}
\begin{abstract}
Transfer reactions are a powerful probe of the properties of atomic nuclei.  When used in inverse kinematics with radioactive ion beams they can provide detailed information on the structure of exotic nuclei and can inform nucleosynthesis calculations.  There are a number of groups around the world who use these reactions, usually with particle detection in large silicon arrays.  Sometimes these arrays are coupled to gamma-ray detectors, and occasionally smaller arrays of silicon detectors are mounted within a solenoid magnet.  Modern techniques using transfer reactions in inverse kinematics are covered, with specific examples, many from measurements made with beams from the Holifield Radioactive Ion Beam Facility at Oak Ridge National Laboratory.  
\end{abstract}

\maketitle

\normalsize

\section{Introduction}
As low-energy radioactive beams become more available there are ever increasing opportunities for exploring the evolution of nuclear structure away from the valley of stability.  In particular, the chances to study nuclei with exotic behaviour, or nuclei that are relevant to astrophysical processes, become tantalizing.  Neutron halo nuclei \cite{Tan85, Alk96, Jen04}  have a large imbalance in the number of protons to the number of neutrons leading to a diffuse neutron tail that surrounds a more tightly bound core.  They are found on, or close to, the neutron drip line where the binding energy for the last neutron or pair of neutrons becomes very small, compared to the typical 8~MeV/nucleon for stable nuclei.  

Higher up in the chart of the nuclides, starting in the neutron-rich nickel region and passing through progressively heavier short-lived nuclei lie the isotopes of the astrophysical rapid neutron capture process, commonly called the r-process \cite{Bur57}.  Elements heavier than iron cannot be produced by fusion owing to the large Coulomb repulsion, which makes such reactions endothermic.  Instead these elements are formed by successive neutron captures and beta decays.  About half of the elements heavier than iron were produced via the r-process, which may occur in the hot, dense neutron-rich environment of a core-collapse supernova.  The path of the r-process is determined by the competition between the neutron-capture Q-value, that is the amount of energy required for a particular nucleus to capture an extra neutron, and its $\beta$ decay half life. Both quantities are strongly influenced by nuclear shell structure, resulting in a zig-zag path through the chart of the nuclides with large discontinuities at the shell closures.  Studying how shell structure evolves away from stability is therefore essential to understanding heavy element synthesis.

\begin{figure}
\includegraphics[width=12cm]{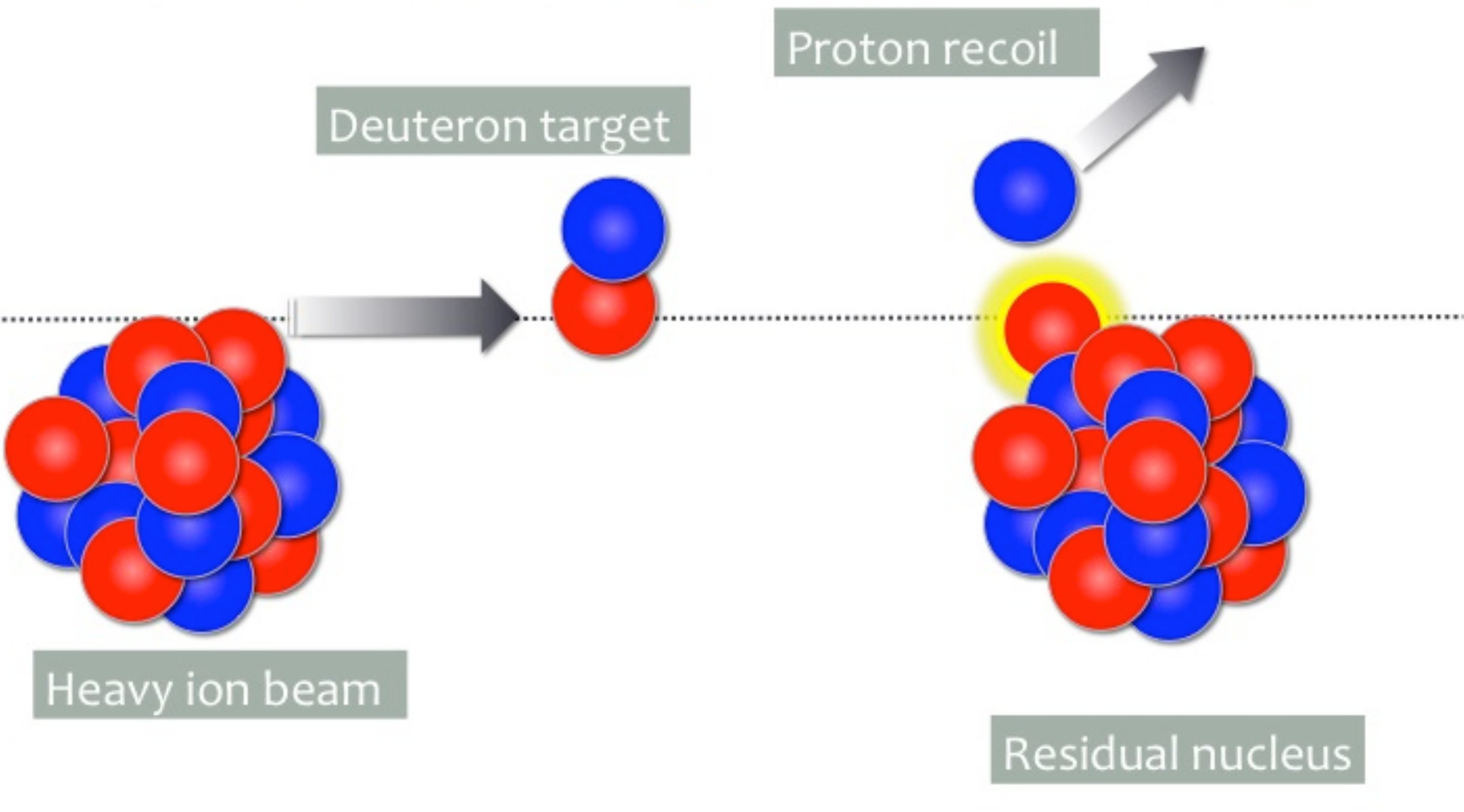}
\centering
\caption{Cartoon of a (d,p) reaction in inverse kinematics.  A heavy ion beam impinges on a deuteron in the target,  resulting in a residual nucleus and a proton. \protect}
\label{inverse}
\end{figure}

Transfer reactions have been used for decades to extract spectroscopic information from nuclei.  Before the early 1990's these reactions were performed in normal kinematics, that is with a heavy target and a light ion beam, with the emergent particles commonly measured at the focal planes of spectrometers.  In this way excellent energy resolution can be achieved, for example a recent measurement using this technique quotes a resolution of 4~keV \cite{Tom11}.  When using a radioactive beam it is necessary to perform the reaction in inverse kinematics \cite{Cat02}, that is with a light target. The focus of this paper will be one-neutron adding reactions, such as (d,p), which are sensitive to the single-particle structure of the residual nucleus.  In the (d,p) reaction, for example, measurements of the angle and energy of protons is related directly to the Q-value of the reaction, and when the beam energy is above the Coulomb barrier, the angular distributions of the protons are indicative of the orbital angular momentum, $\ell$, transfer of the reaction.  Thus, the $\ell$ value (and sometimes by combination with other information the total angular momentum and parity, J$^{\pi}$) of the final state can be extracted.  Additionally, the intensity of the angular distributions is dependent on the overlap between the initial and final states in the reaction.  By performing calculations of the reaction process and scaling\footnote{where the theory is scalable, such as in the Distorted Wave Born Approximation.} to measured differential cross sections (i.e. normalized angular distributions) a quantity referred to as the spectroscopic factor can be extracted.  \\
Transfer reactions in normal kinematics are limited, however, to reactions involving nuclei that can reasonably be incorporated into a target, typically stable nuclei that are solid, or in a solid compound form, at room temperature.  By interchanging the beam and target it is possible to greatly increase the scope of the technique for use with rare ion beams (RIBs).  The inverse kinematics technique (see figure \ref{inverse}) was developed using a $^{132}$Xe beam at the Gesellschaft f{\"u}r Schwerionenforschung (GSI), in Germany \cite{Kra91}.  One purpose of the experiment was to test the (d,p) reaction in inverse kinematics method for future use with a $^{132}$Sn beam; however it is also an example of using these methods to simplify an experiment that would otherwise require a gas target.  This experiment will be described in more detail in section \ref{cp}.

One particular quantity of interest that can be extracted from transfer reactions is the spectroscopic factor, $S$, which is connected to the structure of the nucleus through the single-particle radial overlap function $u_{\ell s j}$ and the normalized wave function, $v_{\ell s j}$:
\begin{equation}
S_{\ell s j}=|{A_{\ell s j}}|^2
\end{equation}
where A$_{\ell s j}$ is the spectroscopic amplitude, and:
\begin{equation}
 u_{\ell s j}(r)= A_{\ell s j} v_{\ell s j}(r)
\end{equation}
see for example \cite{Tho09}.
However, as stated above, as $S$ is extracted from a measured normalized cross section using a calculated angular distribution:
\begin{equation}
S_{exp}=\frac{d\sigma_{exp}/d\Omega}{d\sigma_{calc}/d\Omega}
\end{equation}
$S$ is not an observable of the experiment and, as it relies on the result of a reaction calculation, it is model dependent.  The main sources of uncertainty in the calculations come from the optical (scattering) potentials and the bound state potential of the final state nucleus.  Optical potentials are usually extracted by fitting elastic scattering data and, to a lesser degree, transfer data.  Most of the data existing for both elastic and transfer channels are on stable nuclei, with potentials existing in the literature for individual nuclei (for example \cite{Str77}) and for large regions of the nuclear chart (for example \cite{Var91}).  The latter are referred to as {\it global} optical potentials.  Different optical potentials can produce angular distributions with both different shapes and intensities.  Typically the bound state of the residual nucleus is modelled by a Woods-Saxon potential with the depth adjusted to the binding energy of the state, and the geometry defined by the radius, $r$, and diffuseness $a$.  At center of mass energies close to the Coulomb barrier the intensities of the calculated differential cross sections can be very sensitive to the bound-state potential used.  However, unlike the scattering potentials,  the shape is not dependent on the bound state potential, see for example \cite{Jon11}.  

Since the original GSI measurement, one-neutron transfer reactions in inverse kinematics have been used in a variety of laboratories around the world with beams of long-lived and short-lived nuclei, starting from the pioneering work with a $^{56}$Ni beam by Rehm {\it et al.} \cite{Reh98} (for example \cite{Jon11,Sch12, Win01, Koz06, Jep06, Gau08, Ele09, Kan10, Bac10, Cat10, Tho07, Koz12}).  Generalized methods for studying nuclear structure through one-neutron transfer reactions with RIBs are discussed in the following, with examples of actual experiments.  Most of the examples used are of one-neutron-adding reactions, such as (d,p), commonly termed {\it stripping} reactions, whereas one-neutron-removal transfers, such as (p,d), are commonly called {\it pick-up} reactions.  Methods using the measurement of only the charged recoil (e.g. a proton in the (d,p) reaction) are described in section \ref{cp}. The benefits of measuring $\gamma$ rays in coincidence with charged particles following transfer reactions are described in section \ref{gamma}.   Recent work reviving heavy-ion induced one-neutron stripping reactions, such as ($^9$Be,$^8$Be) and ($^{13}$C,$^{12}$C) is summarized in section \ref{heavy}. As there have been many experiments using stripping reactions in inverse kinematics with RIBs this is not a complete summary of all the experiments performed.  Additionally, there is an emphasis on experiments performed at the Holifield Radioactive Ion Beam Facility (HRIBF) \cite{Bee11} at Oak Ridge National Laboratory in sections \ref{cp} and \ref{heavy}.
	
\section{Measurements with charged particle detectors} \label{cp}
The first experiment to employ a (d,p) reaction in inverse kinematics was the masters thesis experiment of G.~Kraus, with a 5.87~MeV/nucleon beam of $^{132}$Xe at GSI, impinging on deuterated titanium targets backed by 200~$\mu$g/cm$^2$ of aluminium \cite{Kra91}.  Recoil protons were measured in an array of 100 10 x 10~mm$^2$ PIN photodiodes at 375~mm from the target giving an angular resolution of 1.5$^{\circ}$.  A collimator system of 3~-~8~mm slits was used to obtain an energy resolution of around 150~keV FWHM.

\begin{figure}
\includegraphics[width=6cm]{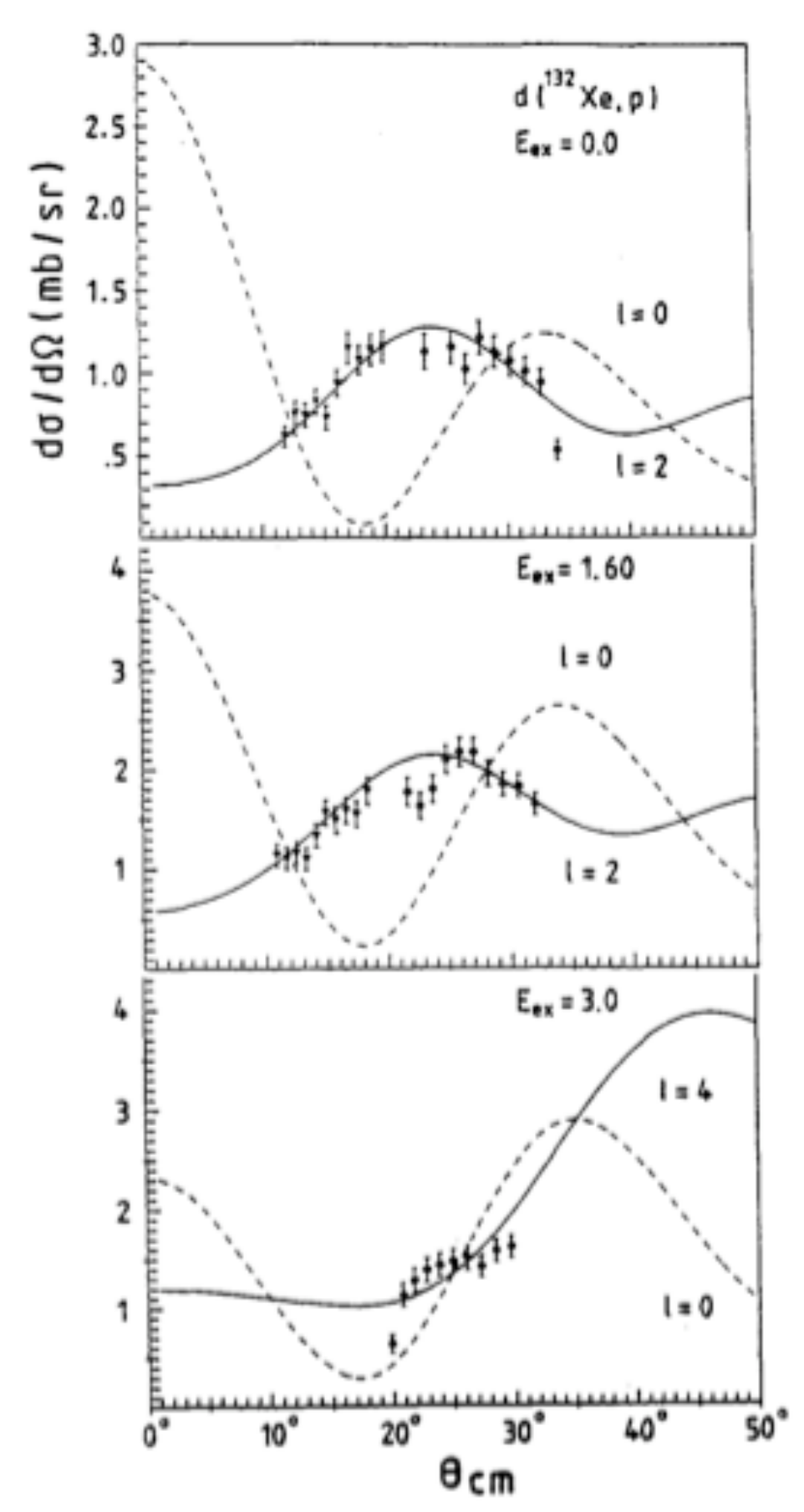}
\centering
\caption{ Angular distributions of protons emitted following the population of the ground, E$_x$~=~1.6~MeV and E$_x$~=~3.0~MeV states in $^{133}$Xe in the reaction d+$^{132}$Xe at 5.87~MeV/nucleon.  The DWBA calculations assume $\ell$~=~0 transfer (dashed) or either $\ell$~=~2 or 4 transfer (solid).  Figure taken from \cite{Kra91}.\protect}
\label{krauss}
\end{figure}

Angular distributions for three states are shown in figure \ref{krauss}, with Distorted Wave Born Approximation (DWBA) calculations, in each case using two different assumptions for the $\ell$ value of the final state (it should be noted that in this reaction and at this beam energy only even-parity states in $^{133}$Xe are expected to be populated).  The calculations, using optical model parameters derived from elastic scattering measured during the experiment, show the sensitivity of the method to the $\ell$ transfer in the reaction; the $\ell$~=~0 curves peak at $\theta_{CM}~=~0^{\circ}$, whereas the $\ell$~=~2 peak is beyond $\theta_{CM}~=~20^{\circ}$, and in the $\ell$~=~4 case, the peak is closer to $\theta_{CM}~=~45^{\circ}$.  This trend is typical for stripping reactions at energies above the Coulomb barrier.

The data are clearly of sufficient quality to differentiate between the various curves allowing the assignment of the $\ell$ value of the states in $^{133}$Xe.  However, it is not possible to directly differentiate between $j$ states using the (d,p) reaction alone unless some preferred alignment is present in the reaction, for example by using a polarized beam or target.  By scaling the calculations to the data, spectroscopic factors for the states populated were extracted with a 40~\% uncertainty.
\begin{figure}
\includegraphics[width=8cm]{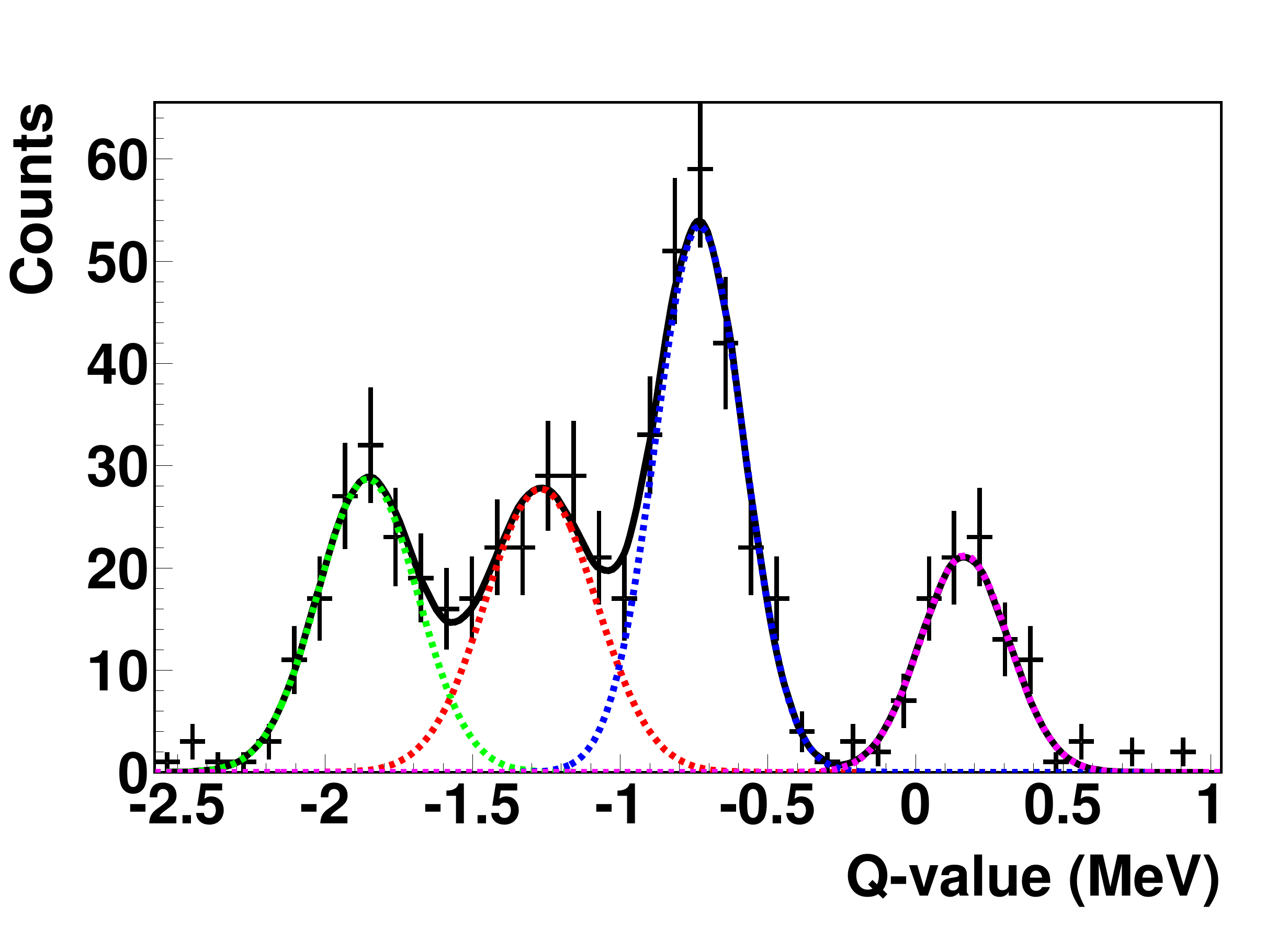}
\centering
\caption{ $Q$-value spectrum for the $^{132}$Sn(d,p)$^{133}$Sn reaction in inverse kinematics at $\theta_{CM}$~=~54$^{\circ}$.  The individual peaks are shown by the coloured dotted lines, whereas the black solid line is a fit to all four peaks, adapted from \cite{Jon10}.\protect}
\label{133_q}
\end{figure}

The $^{132}$Xe measurement was a proof-of-principle study in preparation  for a similar measurement with a $^{132}$Sn beam.  This was realized, some years later, at the HRIBF. A 630~MeV beam of at least 90\% pure $^{132}$Sn bombarded a 160~$\mu$g/cm$^2$ CD$_2$ target at an average intensity of approximately 4 x 10$^4$ $^{132}$Sn ions per second.  Reaction protons were measured in position-sensitive silicon detectors, of the ORRUBA type \cite{Pai07}.  The Q-value of the reaction was calculated on an event-by-event basis from the measured energies and angles of protons emitted in the reaction.  The Q-value spectrum, shown in figure \ref{133_q}, exhibits four peaks, relating to the population of the ground state (purple, right-most peak), and three excited states (from the right, blue, red and green peaks).  The blue peak (third from the right) represents the first observation of a candidate for the p$_{1/2}$ state.  The known energies of the other three states were used as an internal calibration to extract the energy of the previously unknown state, found to have E$_x$~=~1363~$\pm$~31~keV.
\begin{figure}
\includegraphics[width=8cm]{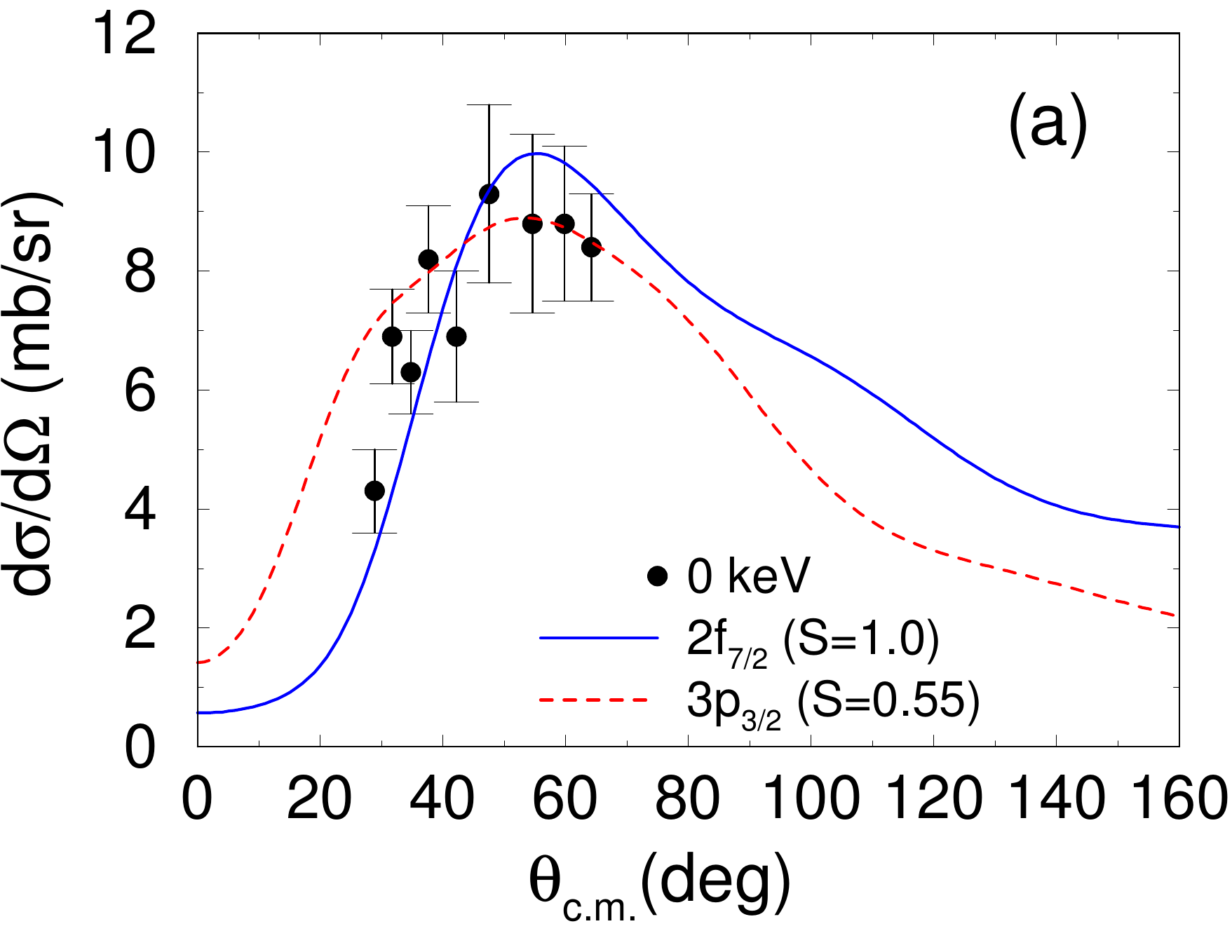}
\includegraphics[width=8cm]{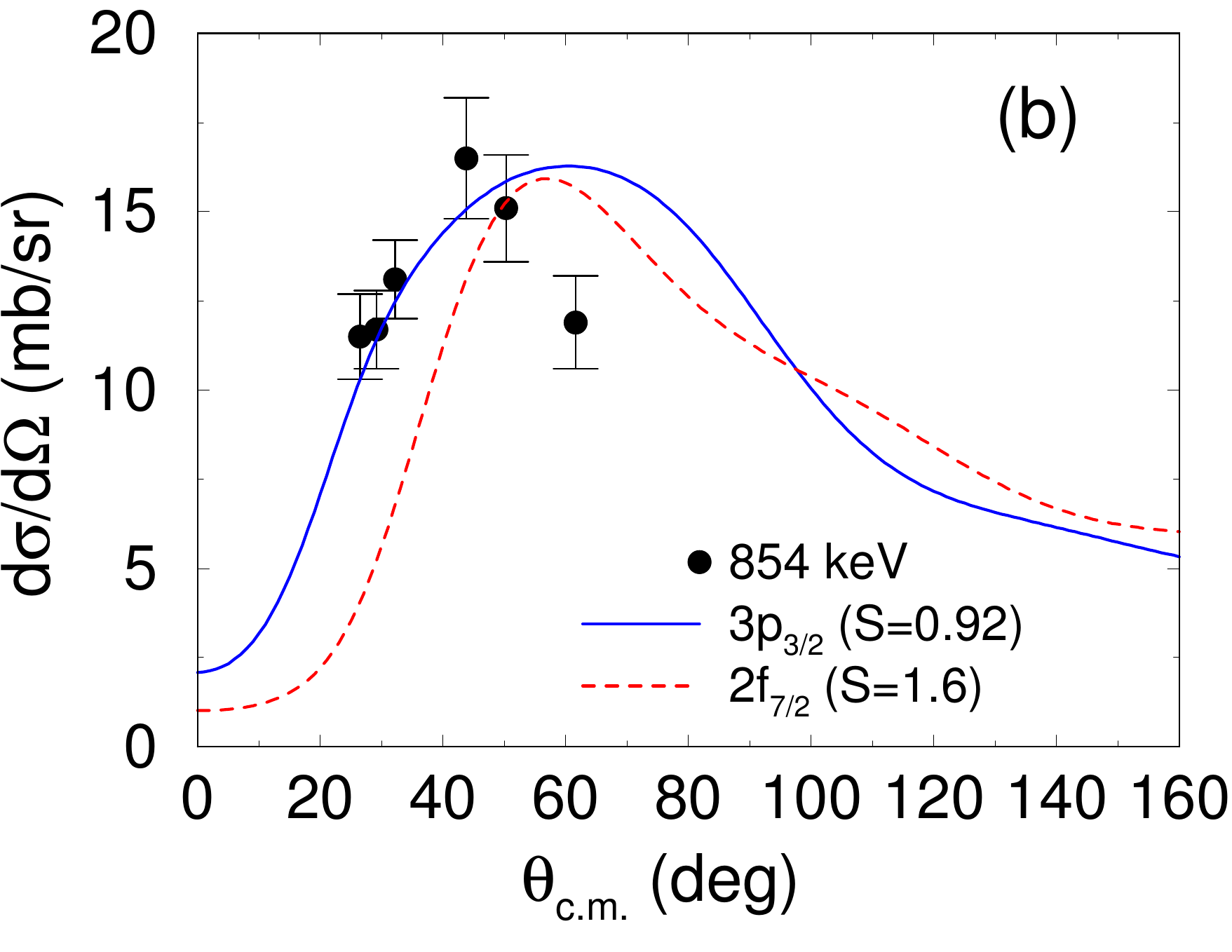}
\centering
\caption{Angular distributions of protons emerging from the $^{132}$Sn(d,p) reaction in inverse kinematics populating (a) the ground state and (b) the state at E$_x$~=~854~keV in $^{133}$Sn.   The solid (blue) curves show the ADWA calculations using the CH89 optical potentials (CH89) \cite{Var91} for the $n\ell j$ for the state, whereas the dashed (red) curves are for the nearest alternate $f$- or $p$-wave single-neutron state.  Adapted from \cite{Jon11}.\protect}
\label{133_ang}
\end{figure}

The data for the population of the two lowest states in $^{133}$Sn, i.e. those relating to the highest energy protons, covered a broad enough range of angles that angular distributions could be plotted.  Figure \ref{133_ang} shows the angular distributions of protons from the population of the ground and 854-keV states, and adiabatic wave approximation (ADWA) calculations assuming an $\ell$~=~1 or $\ell$~=~3 transfer.  These calculations, using the finite range version of ADWA \cite{Joh74, Ngu10}, go beyond standard DWBA by explicitly including deuteron breakup.  The ADWA deuteron adiabatic wave was constructed from the proton and neutron optical potentials from Chapel Hill (CH89) \cite{Var91}, and the calculations were performed using {\sc fresco} \cite{Tho88}.  The calculations were scaled to the data to extract spectroscopic factors for the two states under each assumption.  The ground state angular distribution favours the $\ell$~=~3 transfer, whereas the 854-keV state favours the $\ell$~=~1 transfer, indeed assuming a 2f$_{7/2}$ assignment for this state gives an unrealistically high value for the spectroscopic factor, well outside of the uncertainties.  Spectroscopic factors were also extracted from angle-integrated cross sections for the 1363- and 2005-keV states, and all agree with one within the stated uncertainties (see table \ref{tab1}).  

The spectroscopic factors for $^{133}$Sn were compared to those for $^{209}$Pb which has one neutron more than the benchmark doubly-magic nucleus $^{208}$Pb.  The $^{209}$Pb spectroscopic factors were extracted in exactly the same way, using data from \cite{Ell69} and the same scattering and bound-state potentials as in the $^{133}$Sn case.  The comparison in ref \cite{Jon10} shows that both nuclei have $S$ consistent with 1 for the low-lying states; however in the lead case the values reduce for the higher excited states.  This is not the case for $^{133}$Sn, showing that it is an exceedingly good example of a doubly closed-shell nucleus.  This is important both from the nuclear structure point of view and for calculating the properties of states in even more exotic nuclei important to the r-process.
\begin {table}[t]
\caption{\label{tab1}
Spectroscopic factors of the four single-particle states populated by the $^{132}$Sn(d,p)$^{132}$Sn reaction extracted using the ADWA formalism.  Quoted error margins include experimental uncertainties and 15\% for uncertainties related to the optical potentials used.  Adapted from \cite{Jon11}.\protect \\  }
\begin{tabular}{ccc}
\br
\centering
\textrm{E$ _x $(keV)}&
\textrm {n$\ell$j} &
\textrm{ADWA-CH} \\
\mr
0 			& 2f7/2	           	& $1.00 \pm0.17$ 	\\

854 			& 3p3/2 			& $0.92 \pm0.16$	\\

1363$\pm31$ 	& (3p1/2) 			& $1.1 \pm0.3$	\\

2005 		& (2f5/2)			&$1.2 \pm0.2$	\\
\br
\end{tabular}
\centering
\end{table}

Individual neutron capture rates can influence the final r-process abundance pattern during the freeze-out epoch, when the number of free neutrons has been depleted \cite{Sur01}.   The information extracted from transfer reactions can be used to calculate direct neutron capture (DC), as was done for $^{82}$Ge(n,$\gamma$)$^{83}$Ge using the $^{82}$Ge(d,p)$^{83}$Ge measurement in inverse kinematics \cite{Tho07, Tho05}.   Beun {\it et al.} showed that the neutron capture cross section for $^{130}$Sn in particular can influence the global abundance pattern \cite{Beu09}.  The most important single-particle states for direct capture on $^{130}$Sn are the p$_{1/2}$ and p$_{3/2}$ neutron states, which are populated via s-wave DC followed by the emission of an E1 $\gamma$ ray. These $\ell = 1$ states were observed for the first time in the $^{130}$Sn(d,p)$^{131}$Sn reaction in inverse kinematics \cite{Koz12}, allowing the direct-semidirect capture to be calculated without requiring inputs from mass models, and thereby reducing the uncertainties by orders of magnitude.

Within uncertainties, the spectroscopic factors extracted for $^{133}$Sn using the ADWA formalism \cite{Jon11} agreed with those using DWBA, as presented in \cite{Jon10}.  An example where ADWA calculations gave significantly lower spectroscopic factors than those extracted from DWBA can be seen in the results of the $^{10}$Be~+~d measurements at the HRIBF \cite{Sch12}.  The nucleus $^{11}$Be is unusual in that it has two bound halo states.  Residing relatively close to stability it has been measured numerous times using different techniques including transfer reactions \cite{Aut70, Zwi79, Liu90,For99}, nuclear and Coulomb breakup \cite{Pal03, Fuk04, Lim07}, as well as neutron knockout \cite{Aum00}.  However, the extracted spectroscopic factors for the ground state from these measurements span the region roughly from 0.4 to 0.9.  The major discrepancies come from a nuclear breakup measurement \cite{Lim07} and reanalysis of the (d,p) data \cite{Tim99}.  For the first excited state the situation is worse as only the (d,p) measurements give any information on this state and both have large error bars.  

\begin{figure}[h]
\includegraphics[width=8cm]{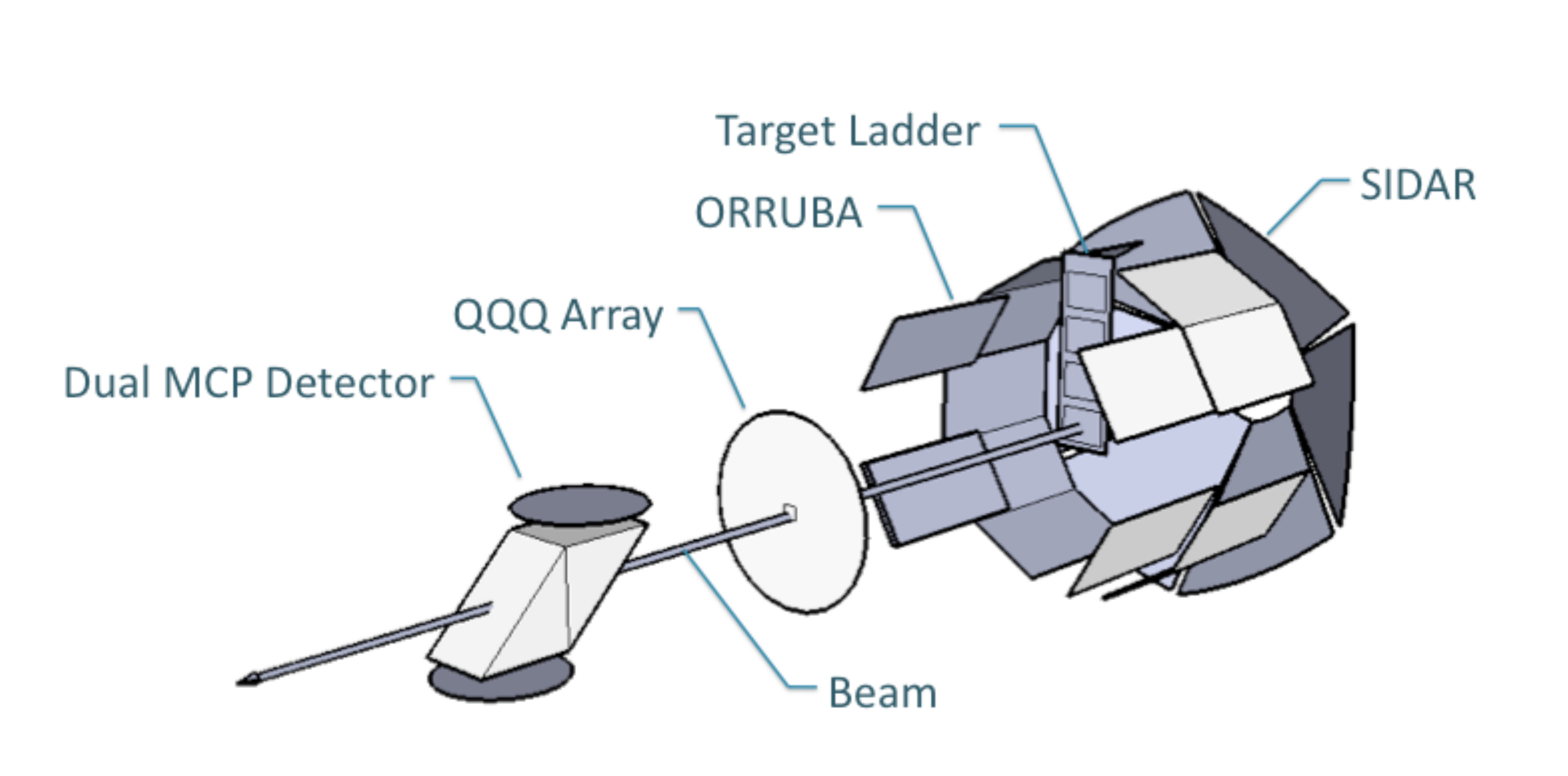}
\centering
\caption{ Detector setup used in the $^{10}$Be~+~d measurement at the HRIBF \cite{Sch12}.\protect}
\label{setup}
\end{figure}

The goal in the work by Schmitt {\it et al.} \cite{Sch12} was to investigate possible sources of inconsistencies from both the experimental and theoretical points of view.  Four measurements were made using primary $^{10}$Be beams with energies ranging from 60 to 107~MeV and having essentially the same setup (a different method for beam counting was employed for the highest beam energy).  Protons were detected in two types of silicon detectors, SIDAR \cite{Bar01} at the most backward angles and ORRUBA \cite{Pai07} around 90$^{\circ}$, see Figure \ref{setup}.  The beam intensity measurement was made using either a dual multichannel plate (107~MeV measurement) or a fast ionization chamber, allowing the differential cross sections to be measured.  Deuteron elastic scattering data were taken simultaneously with the transfer data and were used to inform the choice of optical potentials in the DWBA analysis.

\begin{figure}
\includegraphics[width=6cm]{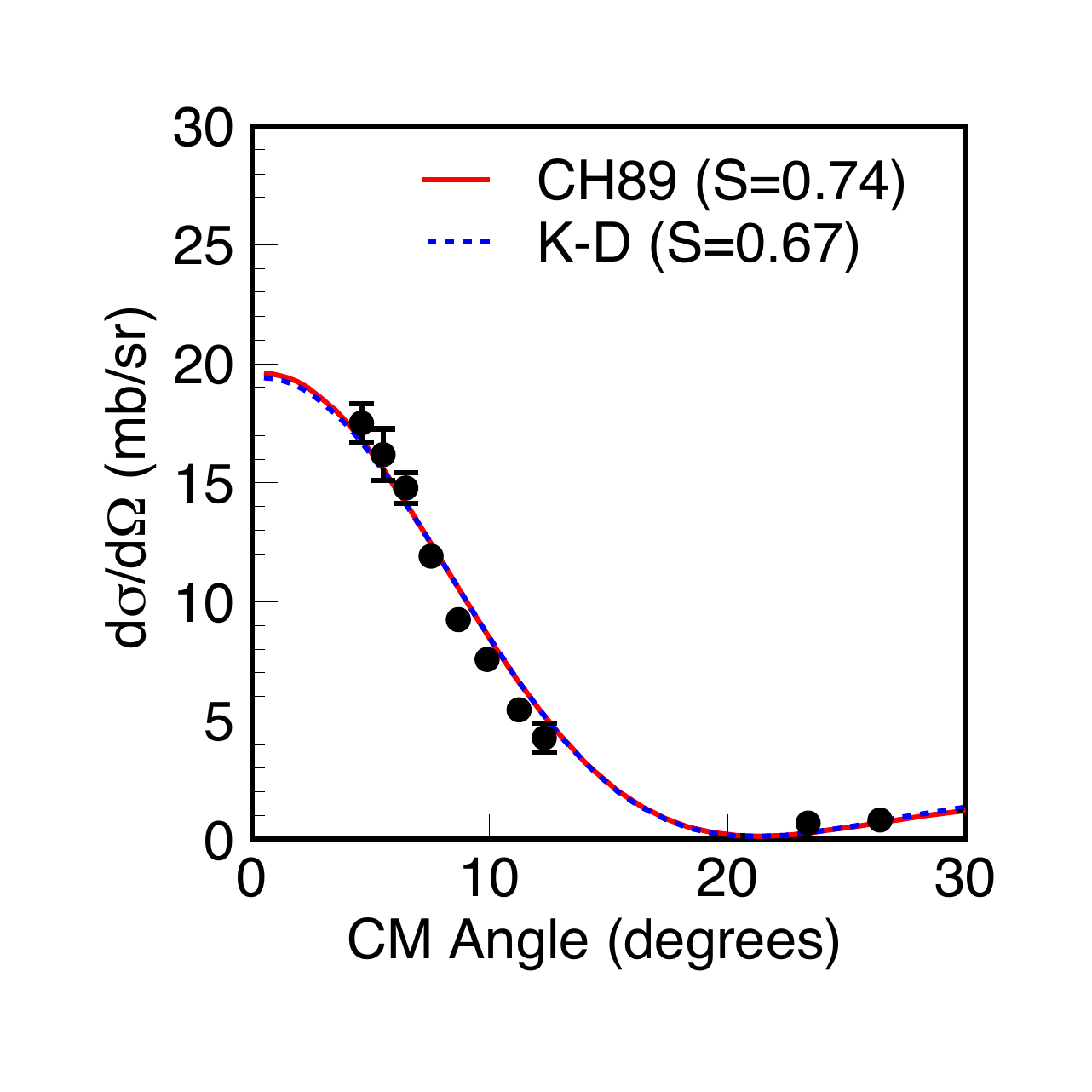}
\centering
\caption{ Angular distribution of protons emitted from the ground state population of $^{11}$Be in the $^{10}$Be~+~d reaction at 107~MeV \cite{Sch12}.  The curves are from calculations using the finite range version of ADWA with optical potentials from CH89 \cite{Var91} (solid, red) and K-D \cite{Kon03} (dashed, blue).   Adapted from \cite{Sch12}. \protect}
\label{11Be_ang}
\end{figure}

The angular distribution for transfer to the ground state at a beam energy of 107~MeV is shown in Figure \ref{11Be_ang}.  The curves shown are for calculations using the finite range version of  ADWA (similar to that used in the $^{132}$Sn~+~d case above) with optical potentials from CH89 \cite{Var91} and A.~Koning and J.~Delaroche (KD) \cite{Kon03}.  The shapes of the calculated curves agree well with the data and there is an approximate 10~\% difference in the spectroscopic factor depending on the potential chosen.  Calculations were also made using a DWBA approach.  The sensitivity of the calculations to the deuteron potential and  the proton potential in the DWBA case was investigated by making calculations varying each in turn.  

\begin{figure}
\includegraphics[width=4.5cm]{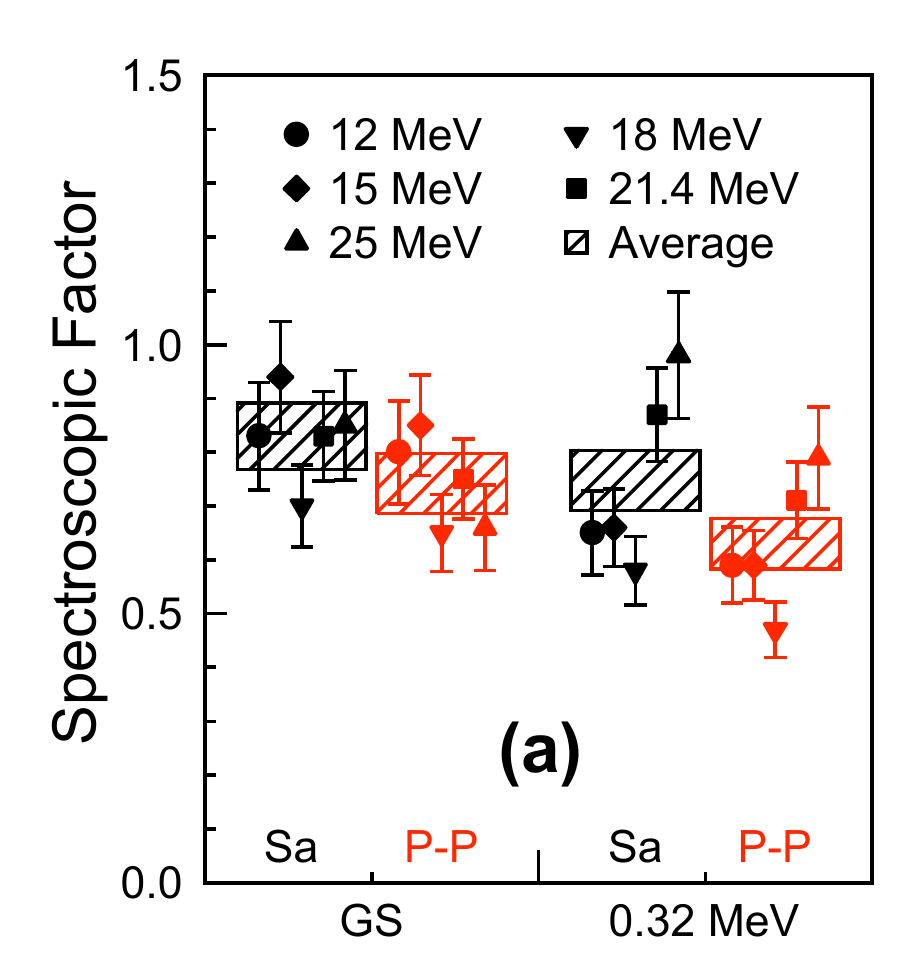}
\includegraphics[width=3.9cm]{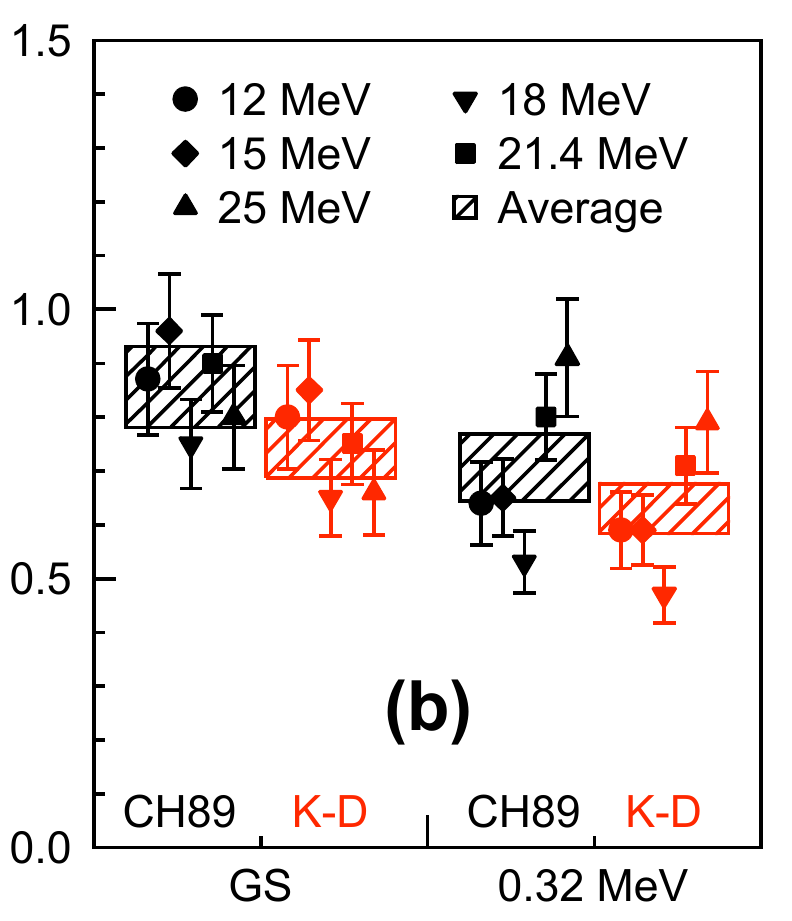}
\includegraphics[width=4cm]{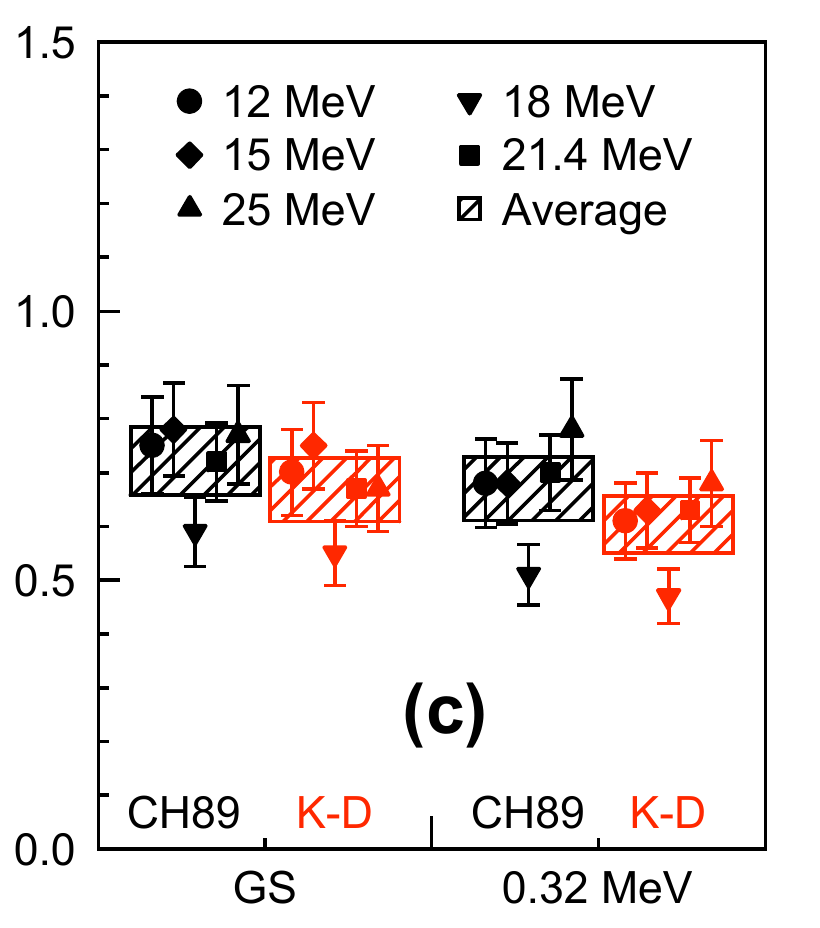}
\caption{ Spectroscopic factors extracted from data taken at the HRIBF \cite{Sch12} and earlier data of Zwieglinski {\it el al} \cite{Zwi79} for both of the bound states, using the a) and b) DWBA and c) the ADWA formalism.  For DWBA calculations there are two sets of optical potentials used, and two different choices were used for the deuteron (in panel a) and for the proton (in panel b).  The box is centered on the average and covers $\pm$ 1~$\sigma$ (see text). Adapted from \cite{Sch12}. \protect}
\label{11Be_SF}
\end{figure}

The results of the spectroscopic factor analysis are shown in figure \ref{11Be_SF} for the measurements in \cite{Sch12} and using the data from \cite{Zwi79} to extract values in the same way at a beam energy of 120~MeV \footnote{It should be noted that the data from \cite{Zwi79} did not agree in shape well with the calculations and hence there is a larger fitting uncertainty associated with those values.  The error bars in figure \ref{11Be_SF} are for experimental uncertainties only.}.  The boxes are centered on the average of the extracted values and show the $\pm 1~\sigma$ ranges.  Two deuteron optical potentials,  Satchler (Sa) \cite{Sat66} and Perey and Perey (P-P) \cite{Per63} were selected using the deuteron elastic scattering angular distributions.  For the proton optical potentials, the choices were the same as in the ADWA analysis, CH89 and KD.
Inspection of the three panels of figure \ref{11Be_SF} shows that the ADWA analysis favours lower values for the spectroscopic factors than those obtained using DWBA.  Also, there is less overlap between the error boxes in the DWBA cases, especially for the excited state, than those from ADWA.  This means that there is greater sensitivity to the optical potentials in the DWBA analysis of this reaction at this range of energies.  The scatter in the data points for the first excited state is also reduced by using ADWA, so the extraction of the spectroscopic factor appears to be less sensitive to the beam energy than in the DWBA case.  However, there is a spuriously low point for the E$_d$~=~18~MeV (E$_{beam}$~=~90~MeV) measurement in every case.  It is not possible to say whether this is simply owing to poorer normalization of this measurement, or if there is some physical effect occurring at this energy which is not taken into account in the calculations.

By consistently measuring the $^{10}$Be~+~d reaction at four different energies and also analyzing the data in the same way for all cases, it was possible to show that including deuteron break-up in the reaction theory and using nucleonic potentials via the ADWA method produced a more reliable extraction of the spectroscopic factors than by using DWBA.  This is different to what was observed in the $^{132}$Sn~+~d case discussed above.  Whether this is an effect of the continuum structure in $^{11}$Be owing to the lower binding energies in this case, it is not possible to say. The $^{11}$Be spectroscopic factors extracted from the average of the four HRIBF measurements are 0.71(5) for the ground state and 0.62(4) for the first excited state, where the quoted uncertainties are solely from experimental considerations.

Importantly, elastic, inelastic and transfer channels were measured simultaneously, and at the same four energies, in a consistent manner.  The elastic channel was used to constrain the choice of optical potential in the transfer analysis.  A more complete study of the $^{10}$Be~+~d reaction at the same energies as in the HRIBF measurement and using a continuum discretized coupled channels (CDCC) formalism is close to completion \cite{Bey12}.

A different method for measuring charged particles emitted from transfer reactions was recently introduced at the ATLAS facility of Argonne National Laboratory.  HELIOS \cite{Wuo07} uses a large-bore superconducting solenoid magnet.  An array of silicon detectors is placed around the magnetic axis of the solenoid, which is collinear with the beam axis.  The cyclotron period (T$_{cyc}$), the time taken for a charged particle to return to the magnetic axis, is dependent on the mass to charge ratio, and independent of energy.  By measuring the position $z$ where the particle returned to the axis, T$_{cyc}$, and the particle energy in the laboratory frame ($E_{lab}$), it is possible to calculate the energy in the centre-of-mass frame using the relationship:
\begin{equation}
E_{lab}=E_{cm}-\frac{1}{2}mv^2_{cm}+\left(\frac{mV_{cm}}{T_{cyc}}\right)z
\end{equation}

A number of experiments have been performed using light RIBs produced via the inflight method, such as \cite{ Bac10, Wuo10, Hof12}  with beams of $^{12}$B, $^{15}$C and $^{19}$O of energies around 6 to 8 MeV/nucleon yielding resolutions ranging from 100 to 175~keV FWHM. In addition, a measurement with a stable $^{136}$Xe beam at 10~MeV/nucleon and targets ranging in thickness between 125 and 175 $\mu$g/cm$^2$ resulted in a Q-value resolution of 90 to 130~keV \cite{Kay11}.

Whether measuring emergent protons from stripping reactions in a large silicon detector, or smaller detectors in a solenoid, the resolution in Q-value will ultimately be limited by the beam quality (an important issues with RIBs) and the energy loss in the target (which becomes less important at higher beam energies).  In order to achieve significantly better energy resolution, meaning $\Delta$E$_x$ of a few keV, it is necessary to measure $\gamma$ rays in coincidence.

\section{Particle-gamma coincidences}\label{gamma}

Recently a number of experiments have been built that detect both charged particles and gamma rays emerging from transfer (and other) reactions.  These include TIARA-Exogam \cite{Lab10,Sim00} at GANIL, TREX-MINIBALL \cite{Bil07,Ebe01} at REX-ISOLDE and SHARC-TIGRESS \cite{Dig11,Sve05} at ISAC II.

Typically, the resolution in the excitation energy (or Q-value) is compromised owing to the need to place silicon detectors close to the target, directly impacting the angular resolution. Additionally, owing to the lower efficiencies associated with gamma ray detection over charged-particle detection thicker targets are required, especially when the beams are weak.   This further compromises the charged particle resolution. However, this loss in charged-particle resolution is more than compensated for by the measurement of $\gamma$ rays in coincidence.  Not only do the $\gamma$ rays provide an improvement in resolution of around an order of magnitude or more over measuring charged particles alone, they also give extra information that is otherwise inaccessible, such as observing cascades of $\gamma$ rays through states that were not populated in the reaction.  

\begin{figure}[ht]
\begin{minipage}[b]{0.7 \linewidth}
\centering 
\includegraphics[width=\textwidth]{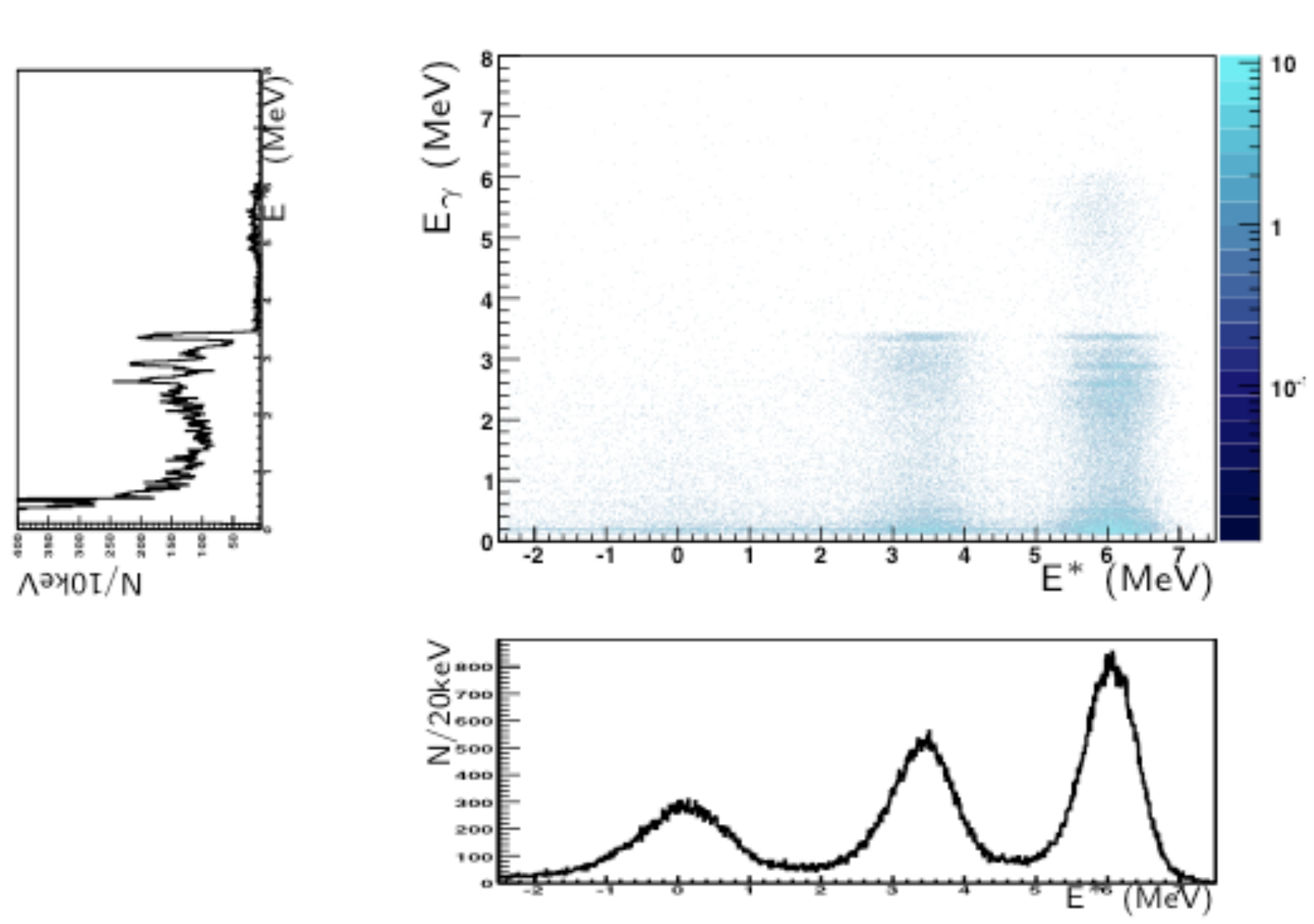}
\end{minipage}
\begin{minipage}[b]{0.25\linewidth}
\centering
\includegraphics[width=\textwidth]{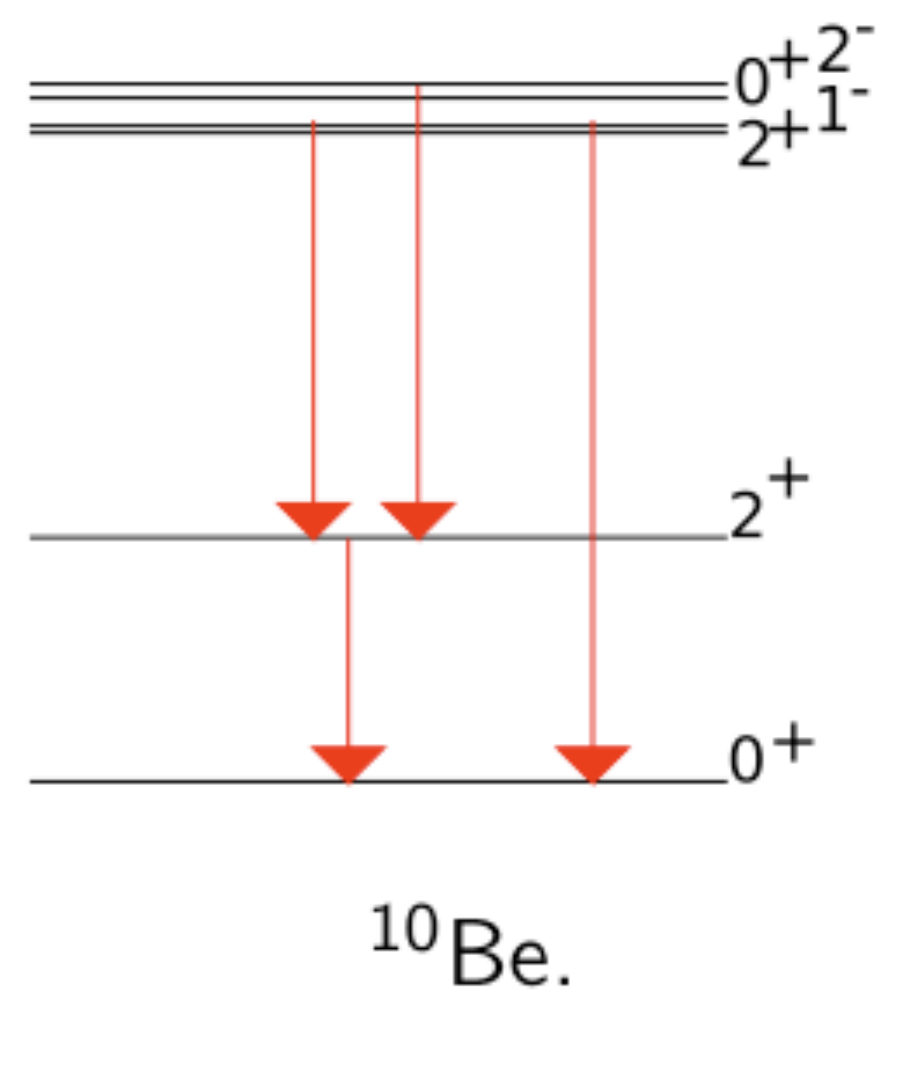}
\end{minipage}
\caption{ Energy of $\gamma$ ray versus excitation energy $E^*$ of states populated in $^{10}$Be via the $^{11}$Be(d,t) reaction in inverse kinematics.  The projections onto the E$_{\gamma}$ and E$^*$ axes are shown.  A schematic of the $^{10}$Be level scheme and observed $\gamma$ rays is shown on the right.  Taken from \cite{Joh11}. \protect}
\label{10Be}
\end{figure}

A recent set of experiments was performed with a 30.7~MeV $^{11}$Be beam, produced at REX-ISOLDE, incident on deuterons in a CD$_2$ target with particles detected in TREX \cite{Bil07} and $\gamma$ rays measured in MINIBALL \cite{Joh11}.  The one-neutron pickup, elastic and one-neutron stripping channels were identified via the emitted proton, deuteron or triton to study states in $^{10}$Be, $^{11}$Be and $^{12}$Be respectively. 

 In figure \ref{10Be} the $\gamma$ ray energy, E$_{\gamma}$, is plotted as a function of excitation energy, E$^*$, extracted from the energy and angle of the triton emerging from the $^{11}$Be(d,t)$^{10}$Be reaction in inverse kinematics.  Points located along the diagonal, where E$_{\gamma}$~=~E$^*$, represent events where there was direct  deexcitation of the populated state via a $\gamma$ ray to the ground state.  For example, there is a group of events at around E$_{\gamma}$~=3.4~MeV that are correlated with the population of the 2$^+$ state at 3.368~MeV.  Similarly at around 6~MeV in both E$_{\gamma}$ and E$^*$ a more diffuse set of events is present representing the direct transition from the group of states around 6~MeV to the ground state.  Additionally, some of the deexcitation of those states goes through the 2$^+$ state.  In this way, the charged-particle measurement gives information on the state populated in the reaction, whereas the $\gamma$ rays give the precise energy of the state, and information on the decay out of the state.
 
\begin{figure}
\centering 
\includegraphics[width=8cm]{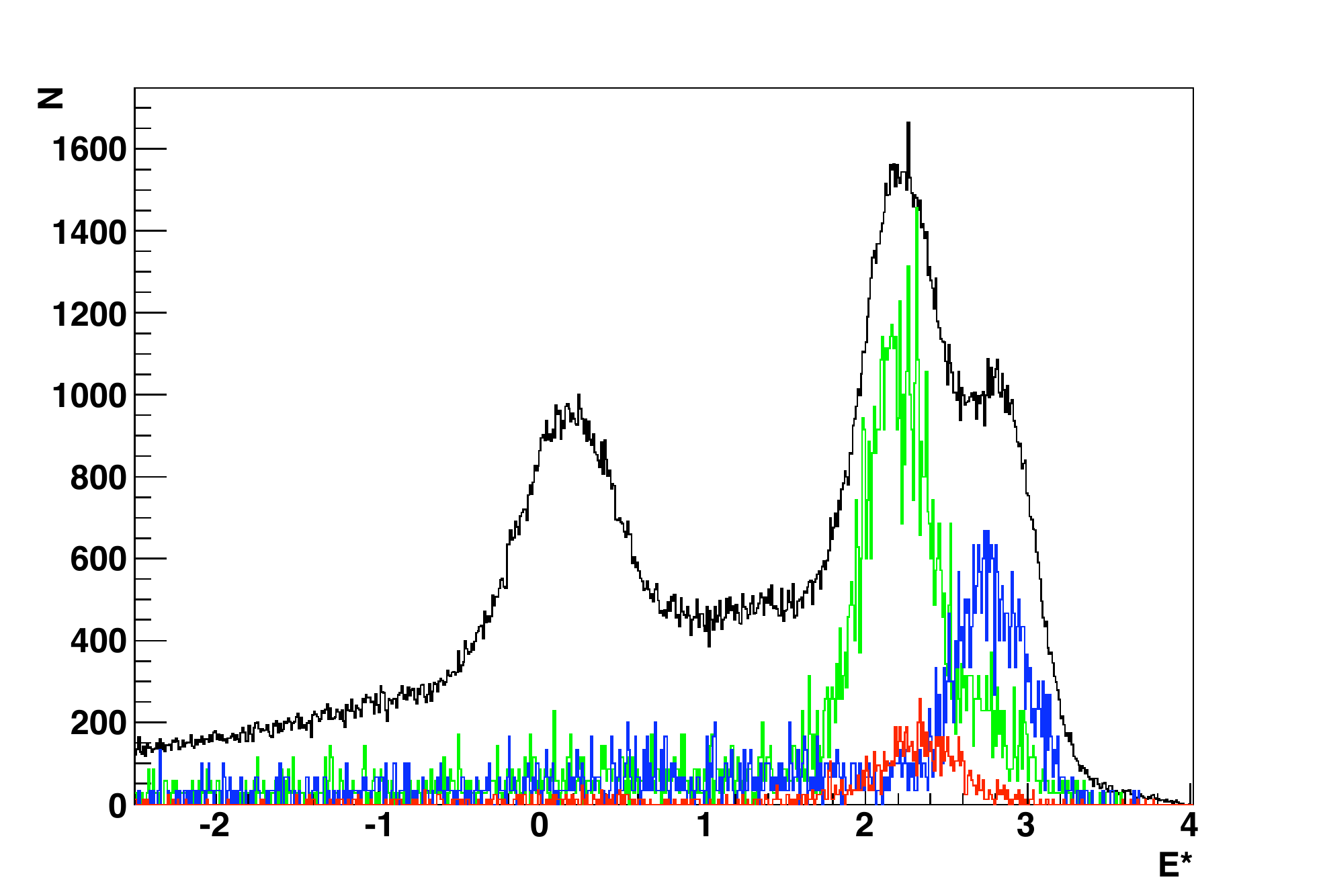}
\includegraphics[width=3.5cm]{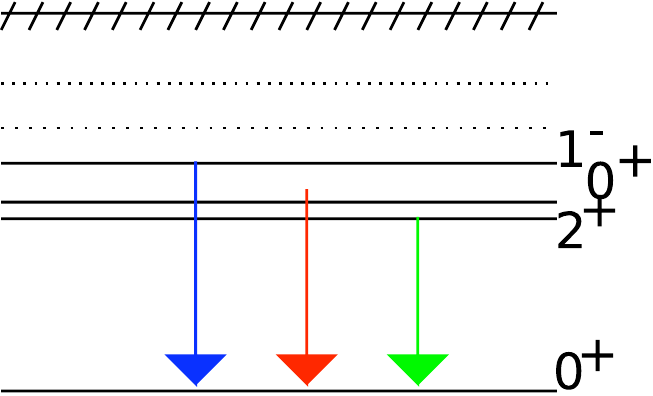}
\caption{ Excitation energy, $E^*$, of states populated in $^{12}$Be via the $^{11}$Be~+~d reaction (black) with those components that come in coincidence with different energy $\gamma$ rays shown as individual curves.  The curve peaking at $E^* \approx$ 2.1~MeV (green) is gated on the  $2^+\rightarrow 0^+$ decay, and the curve peaking at $E^*\approx 2.7$~MeV (blue) is gated on the $1^-\rightarrow 0^+$ decay, as shown on the level scheme on the right. The smaller (red) line is in coincidence with the $0^+\rightarrow 0^+$ decay, which is an E0 transition observed by two-photon decay. Taken from \cite{Joh11}.\protect}
\label{12Be}
\centering
\end{figure}

Another way of combining the charged-particle and $\gamma$ ray information is shown in figure \ref{12Be} for the (d,p) channel going to $^{12}$Be.  The black histogram shows the total excitation energy, whereas the data shown in colour represent the excitation energy profile gated on individual $\gamma$ rays, shown in the level scheme.  Demanding a $\gamma$ ray in coincidence eliminates both the data from population of the ground state and the vast majority of the background.  The highest energy peak is mostly from the 1$^-$ state; however, the three excited states populated are not well resolved.  Without the $\gamma$ ray measured in coincidence, it would not be possible to see the contribution from the second excited state.

Complex transfer experiments with RIBs at GANIL have been performed with combinations of the charged-particle detection arrays TIARA and MUST 2 \cite{Pol05}, the $\gamma$ ray array EXOGAM and the spectrometer VAMOS, such as \cite{Ram09}.  The detection of the heavy recoiling nucleus in VAMOS can be used to tag the reaction channel, greatly reducing the background from reactions on beam contaminants and non-time-correlated events, and in some cases, the angle of the recoil can be used to improve the Doppler correction of the $\gamma$ ray measurement.

The ideal situation for measuring transfer reaction with $\gamma$ rays is where the silicon detectors measuring the charged particles are at a great enough distance from the target that angular resolution is not the limiting factor.  This requires that the gamma detection array is large enough to cover the total solid angle at a large ($> 10$~cm) radius.  Such a system is being planned for using ORRUBA \cite{Pai07} with GAMMASPHERE \cite{Lee90} at ATLAS.

\section{Heavy-ion induced reactions}\label{heavy}

Until this point we have considered transfer reactions induced by light nuclei, meaning protons or deuterons, on rare ion beams.  However, there are advantages in some cases in using heavier nuclei, such as $^9$Be or $^{13}$C, to induce the reaction.  Although there are examples where the charged ions alone were measured (e.g. \cite{Ash03}), the focus here is on a different technique where the charged particles are used to identify the reaction channel and then the $\gamma$ rays coming in coincidence are analyzed.  

\begin{figure}
\centering 
\includegraphics[width=8cm]{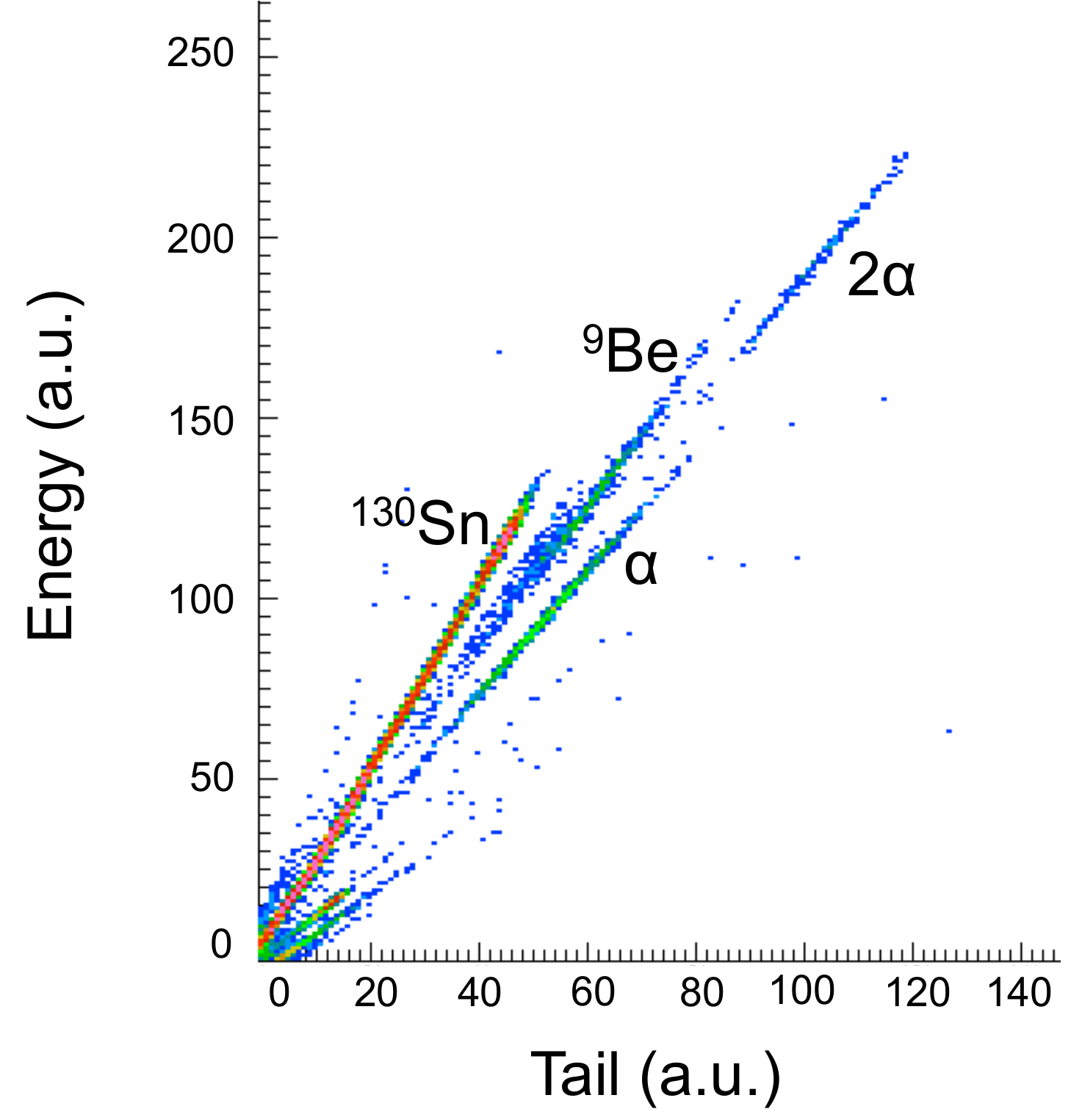}
\caption{Particle identification in a ring 1 detector of HyBall at $\theta_{lab}=10.5^\circ$ following the $^{130}$Sn+$^9$Be reaction at the HRIBF.  The 2$\alpha$ group is the cleanest indication of a one-neutron transfer reaction.  The $^{130}$Sn and $^9$Be groups are from the elastic scattering of the beam and the target nuclei respectively.  Taken from \cite{130Sn}.\protect}
\label{PID}
\centering
\end{figure}

The ($^9$Be,$^8$Be) reaction favours low angular momentum transfer owing to the weak binding of $^9$Be ($S_n$~=~1.66~MeV) which is even smaller than that for the deuteron ($S_n$~=~2.2~MeV) \cite{Sta77}.   However, the weak binding also means that the $^9$Be is likely to break up in the reaction, making it difficult to analyze under a simple DWBA formalism.   Additionally, $^9$Be itself is strongly deformed, and has a strong cluster nature ($\alpha + \alpha + n$ or $^5He + \alpha$) which complicates the reaction theory analysis  \cite{Kee05}.   The advantage of the $^9$Be reaction is that the recoiling $^8$Be is unbound and breaks up into two $\alpha$ particles, giving a unique signature that provides a clean gate for looking at $\gamma$ rays in the recoiling nucleus (see \cite{Gun01,Rad05}), as shown in Figure \ref{PID}.   

The more commonly-used heavy-ion induced reaction is ($^{13}$C,$^{12}$C) \cite{Bon74}.  In this case, the stronger binding of the last neutron in $^{13}$C ($S_n$~=~4.95~MeV) leads to smaller Q-values than in ($^9$Be,$^8$Be), favouring transfer to higher-spin states \cite{Sta77, Bon74}.  As spin-flip transitions are favoured, non-spin-flip transitions are hindered \cite{Bon83, Bri72,Bla04}, and the last neutron in $^{13}$C is in the p$_{1/2}$ orbital, states with $j=\ell+1/2$ are preferentially populated. In the ($^9$Be,$^8$Be) case it is the $j=\ell-1/2$ states that are preferred as the last neutron in $^9$Be is in the p$_{3/2}$ state; however this tendency is tempered somewhat owing to the weak binding of $^9$Be. The ($^{13}$C,$^{12}$C) reaction has been used to extract radii from known or assumed spectroscopic factors (e.g. \cite{Fra79}), Asymptotic Normalization Coefficients (ANC) or relative spectroscopic factors (e.g. \cite{Kay09}), or to populate states to be used in general $\gamma$ ray spectroscopy (e.g. \cite{Ish08}).

An experiment was performed at the HRIBF with a $^{134}$Te beam to study states across the $N=82$ shell gap via the ($^9$Be,$^8$Be) and ($^{13}$C,$^{12}$C) reactions in inverse kinematics \cite{Rad05, All12}.  The charged particles were detected in the forward four rings of HyBall \cite{Gal}, comprising 40 crystals of cesium iodide with $\theta_{lab} = 7^{\circ}-60^{\circ}$ . The CLARION array of 11 high purity germanium detectors \cite{Gro00} was used to detect coincident $\gamma$ rays.  With a beam intensity of approximately $3\times10^5$pps of $^{134}$Te it was possible to obtain clear coincidences in particle-$\gamma$-$\gamma$ spectra allowing the identification of the previously unobserved $i_{13/2}$ state in $^{134}$Te at around 2.11~MeV.  This assignment was supported by angular correlations between the carbon recoil from the $^{134}$Te+$^{12}$C reaction.

Experiments were run during Spring 2012 at the HRIBF to study states in odd-mass neutron-rich tin isotopes using beams of $^{124-132}$Sn, following on from the $^{134}$Te work.  Preliminary analysis has revealed candidate $\gamma$ rays for the depopulation of the four single-particle states in $^{131}$Sn \cite{130Sn} that were revealed in the (d,p) study \cite{Koz12},  high resolution measurements of states in $^{126,128}$Sn also populated in (d,p) \cite{128Sn}, and a candidate for the $i_{13/2}$ state in $^{133}$Sn \cite{132Sn}.

\section{Summary}
Transfer reactions have seen a revival since the mid-1990's when methods for using them in inverse kinematics with rare ion beams were developed.  One-neutron stripping reactions represent a powerful method for extracting spectroscopic information relating to single-particle states.  There has been intense work at a number of RIB facilities around the world to use these techniques to study nuclei both with and without coincidences with $\gamma$ rays.  The most commonly used stripping reaction is (d,p), which is also the most well understood reaction from the theoretical point of view, owing to the simplicity of the light particles involved.  However, in some cases it is important to include the breakup channel for the weakly bound deuteron in the analysis of (d,p) data.

Heavy-ion induced reactions are also undergoing a resurgence.  As  reactions with more strongly bound nuclei, such as $^{13}$C, have different selectivity to those with weakly bound nuclei, such as $^{9}$Be and the deuteron, the combined information from these studies can reveal the $j$ state of the transferred nucleon, without the need for polarized beams or targets.

As new facilities become available around the world, new detection setups for transfer reaction studies are being developed, allowing ever more sensitive studies to be performed.

\subsection{Acknowledgments}
I would like to thank all of my collaborators on the experiments described here, especially James Allmond, Daniel Bardayan, James Beene, Anissa Bey, Jolie Cizewski, Alfredo Galindo-Uribarri, Ray Kozub, Brett Manning, Filomena Nunes, Steven Pain, David Radford, Kyle Schmitt and the ORRUBA-RIBENS, and CLARION-HYBALL collaborations.  Additionally I would like to thank Jacob Johansen, Karsten Riisager and their collaborators for allowing their data to be presented here.
This work was supported by the US Department of Energy
under contract numbers DE-FG02-96ER40983 (UT), DE-SC0001174(UT). \\

\bibliographystyle{unsrt.bst}
\bibliography{Nobel}

\end{document}